\documentclass{ws-ijmpc}

\begin{document}

\markboth{S. Bhowmick and D. Ghosh}
{Engineering synchronization }

\catchline{}{}{}{}{}

\title{Targeting engineering synchronization in chaotic systems}

\author{Sourav K. Bhowmick}

\address{Department of Electronics,\\ Asutosh College, \\Kolkata 700026, India.}

\author{Dibakar Ghosh \footnote{dibakar@isical.ac.in}}

\address{Physics and Applied Mathematics Unit, \\Indian Statistical Institute, \\Kolkata 700108, India.}

\maketitle

\begin{history}
\received{Day Month Year}
\revised{Day Month Year}
\end{history}

\begin{abstract}
A method of targeting engineering synchronization states in two identical and mismatch chaotic systems is explained in details. 
The method is proposed using linear feedback controller coupling for engineering synchronization such as mixed synchronization, 
linear and nonlinear generalized synchronization and targeting fixed point. The general form of coupling design to target any desire 
synchronization state under unidirectional coupling with the help of Lyapunov function stability theory is derived analytically. A scaling 
factor is introduced in the coupling definition to smooth control without any loss of synchrony. Numerical results are done on two mismatch 
Lorenz systems and two identical Sprott oscillators.

\end{abstract}

\ccode{PACS Nos.: 05.45.Xt, 82.40.Bj}

\section{Introduction}
Chaotic system is defined by its complex dynamical behaviors, particularly their extreme
sensibility on initial conditions and parameter variations, which make their behaviors long-term unpredictable. The coupled chaotic system exhibit
interesting complex feature like  riddled basin, on-off intermittency, different types of synchronization and amplitude death between the oscillators. In 1665,
Huygens discovered an odd kind of sympathy in two pendulum clocks suspended on a beam. This was visualized  as an interesting example
of synchronization phenomena between two clocks. Recently in last two decades this phenomenon has  become an important topic to be
discussed in the aspect of nonlinear coupled systems i.e. mainly in chaotic system.

\par Pecora and Carrol\cite{picora} introduced a method to synchronize two identical systems with different initial conditions. Several types of synchronization
like complete synchronization(CS)\cite{pikovsky}, antisynchronization(AS)\cite{danaAS,danaAS2,danaAS3,danaAS4}, lag synchronization(LS)\cite{rose},phase synchronization(PS)\cite{rosenblum},generalized synchronization(GS)\cite{abarbanel,abarbanel2,chenijmpc}, projective synchronization\cite{diba,diba2,diba3}, an alternate approach like engineering synchronization\cite{danaeng,danaeng2,danaeng3,ijmpcdghosh} etc. have been reported before.
 In CS the state variables of
coupled system are completely coincide where as phase difference $\pi$ exist between them for AS. The state variables of the coupled
oscillators emerge into a mixed synchronization (MS)\cite{hasler,yclai,prasad,opcl}, where a pair of state variables develops a CS state while 
another pair is in a AS state. It is true that existence
of a MS state was reported earlier\cite{hasler}  in coupled co-rotating Lorenz systems for a specific scalar coupling. This emergent MS state has justification in
the inherent axial symmetry of the chaotic flow and map\cite{yclai}. Recently MS state is emerges in counter-rotating oscillators under linear diffusive 
coupling\cite{prasad,prasad2,prasad3}. It was recently reported\cite{opcl,opcl2} that, given a definition of a model system, one can design a coupling to target MS state in delayed 
and nondelayed systems. The combination of CS and AS provide a way to cryptographic encoding for digital signal through parameter modulation\cite{plams,dibaps}.

\par The other kind of synchronization namely GS was introduced which tried to explain emergence of a kind of functional relationship between the coupled 
oscillators in drive - response mode.
GS is thus seen as an evolution of a functional relationship between state variables of driver-response system. This functional relationship for GS is 
generally unknown under conventional diffusive coupling.
Another most intriguing effects in dynamical system is so-called amplitude death (AD)\cite{ad,ad2,ad3}, dealing with suppression
of oscillations to a steady state. Amplitude death in coupled systems generally occurs when the amplitude
of the oscillation is damped out to a steady state. This all types of dynamical phenomenon also reported earlier using different coupling for getting desired state. 
For getting desired synchronization state different coupling configuration has been investigated such as unidirectional\cite{uni,uni2}, bidirectional\cite{pikovsky}, 
repulsive\cite{repulsive}, inhibitory\cite{inhibitory,inhibitory2} or excitatory\cite{excitatory}, synaptic\cite{synaptic} coupling etc. To observe different types of 
synchronization state between two or more oscillators, major efforts are given on the strength of coupling between them and parameter mismatch or 
noise intensity. Under linear coupling, in most studies, different synchronization state i.e. CS, AS, PS, LS, GS are observed by varying the coupling 
strength but above critical coupling strength most of cases desynchronization regime observed.
\par Recently the concept of engineering synchronization\cite{danaeng} in nonlinear oscillators given importance, for practical application. In 
engineering synchronization, it is assumed that the definition of chaotic flow is known but the definition of coupling unknown {\it apriori}. 
In this paper we address the question ``how to define the coupling between two coupled oscillators to get desired synchronization state when 
the definition of flow is known?".
\par In this paper, we propose linear feedback controller (LFC) to target engineering synchronization state. We design coupling definitions to target 
engineering mixed synchronization, linear and nonlinear generalized synchronization and target fixed point. Partial symmetry of the system is the necessary 
condition for MS state. Note that this restriction is valid for MS under linear conventional diffusive coupling too. Here, we remove this restriction to get MS state. 
Given a model system and target synchronization state, the coupling is first defined analytically to establish engineering synchronization state between two 
chaotic systems. The stability condition for targeting engineering synchronization state is derived using Lyapunov functional theory. A smooth control from 
one synchronization state to another synchronization state namely CS to AS and vice-versa is achieved by inserting a scaling matrix in the definition of coupling. 
This is the main contrast to what usually observed for conventional linear diffusive coupling. Engineering linear and nonlinear generalized synchronization is 
also observed for a suitable choice of target function. In GS state the functional relationship between drive-response systems is generally unknown under 
conventional diffusive coupling. But targeting engineering generalize synchronization, the functional relationship is known which is a contrast to the 
conventional diffusive coupling. We also target the fixed point which may be a fixed point of the system or any other fixed point whereas the driving 
system is in chaotic state and response system target to a fixed point. Amplification and attenuation of response system's attractor is also possible using 
LFC coupling. The proposed LFC coupling is not dependent on system's parameters. For a given system model and a desired synchronization state, if the 
coupling term designed 
then smooth control of one synchronization state to another synchronization state with amplification or attenuation is possible without loss of synchrony. 
The scaling of a chaotic attractor has potential application in secure communication. Numerical simulations are done using mismatch Lorenz system and 
identical Sprott oscillators. 
\par We organize the paper as follows: Sec. II, general theory of linear feedback controller (LFC) is discussed for targeting engineering synchronization 
between two non-identical systems. The theory is also valid for identical systems taking mismatch term as zero. In Sec.III, the theory for LFC using 
unidirectional coupling is elaborated how to realize MS state i.e. co-existence of CS and AS between different state variables using numerical example 
of non-identical Lorenz system\cite{lorenz}. Numerical example on identical Sprott oscillator\cite{sprott} for engineering linear generalized synchronization 
is discussed in Sec. IVA. Sec. IVB is elaborated with engineering nonlinear generalized synchronization using identical Sprott oscillator. In Sec. V, targeting 
fixed point is discussed on Sprott system where the driving states are in chaotic state and response states are in a desired fixed point state. The results are 
summarized in Sec. VI.   

\section{General theory of coupling}
We consider the driver system as
\begin{equation}
\dot{x}=f(x, \mu)+\bigtriangleup f(x, \mu )
\end{equation}
where $x \in R^n$, $\mu$ is the vector of parameters and $\bigtriangleup f(x, \mu )=f(x, \mu+\delta \mu)-f(x, \mu)$, in general and $\delta \mu$ denotes the 
mismatch parameters. Otherwise if all parameters are appeared in linear form in $f(\cdot )$ then $\bigtriangleup f(x, \mu )=f(x, \delta \mu)$.
\par Now we consider the response system as follows
\begin{equation}
\dot{y}=f(y, \mu) +U
\end{equation}
where $U$ is controller which we want to derive using Lyapunov stability theory. The error signal of the coupled system is defined as
\begin{equation}
e=y-\phi(x)
\end{equation}
where $\phi(x)$ is the target or goal function. Error dynamics is
\begin{equation}
\dot{e}=\dot{y}-D \phi(x) \dot{x}=f(y, \mu)-D \phi(x)[f(x, \mu)+\bigtriangleup  f(x, \mu)]+U
\end{equation}
where $D \phi(x)$ is the Jacobian matrix of the function $\phi(x)$.
We choose the controller $U$ as follows
\begin{equation}
U(t)=u'(t)+w(t)
\end{equation}
where $u'(t)$ and $w(t)$ are the active and linear feedback controller respectively. So the error system to be controlled is now a linear system with the 
control input function $w(t)$ as function of the error states. When the error system will
be stabilized by the feedback $w(t)$, the error will converge to zero as $t\rightarrow \infty $ which implies that the drive and response system are
globally synchronized. To achieve this goal we choose $w(t)$ such that, $$w(t)^T= A.e(t) $$ where $A$ is a matrix of order $n\times n$. 
So $A$ should be chosen properly to get target synchronization state.
Now we consider the Lyapunov function as follows
\begin{equation}
V=\frac{1}{2}e^T e
\end{equation}
where $T$ is transposition of matrix. It can be easily verified that $V(t)$ is a non-negative function. Assuming the system's parameters are all known, 
we make an appropriate choice of the controllers $u'(t)$ and $w(t)$ so that $\frac{dV}{dt}<0.$  The error dynamics will be asymptotically globally 
stable if $\frac{dV}{dt}<0$ and thereby realize any targeted synchronization state.

\section {Targeting Engineering Mixed Synchronization}
Consider the three dimensional autonomous chaotic Lorenz system\cite{lorenz}, described by the set of equations:
\begin{eqnarray}
\label{idlor}
&&\dot{x_1}=\sigma(x_2-x_1) \nonumber\\
&&\dot{x_2}=rx_1-x_2-x_1x_3 \nonumber\\
&&\dot{x_3}=x_1x_2-\beta x_3
\end{eqnarray}
where $x_1, x_2, x_3$ are the state variables and $\sigma, r, \beta$ are parameters of the above system. The system exhibits chaos when
$\sigma=10 $, $\beta=\frac{8}{3} $, $r=28.$

We consider the system $(7)$ as the driving system with parameter mismatch as the following equation $(8)$ where $\delta \sigma ,
\delta r , \delta \beta$ are the parameter mismatch corresponding to $\sigma, r, \beta$ respectively.
\begin{eqnarray}
\label{missrossdir}
&&\dot{x}_1= (\sigma+\delta \sigma)(x_2-x_1)\nonumber\\
&&\dot{x}_2= (r+\delta r)x_1-x_2-x_1x_3\nonumber\\
&&\dot{x}_3=x_1x_2-(\beta +\delta \beta)x_3
\end{eqnarray}

We consider the response system as follows
\begin{eqnarray}
\label{missrossres}
&&\dot{y}_1= \sigma(y_2-y_1)+u_1\nonumber\\
&&\dot{y}_2= ry_1-y_2-y_1y_3+u_2\nonumber\\
&&\dot{y}_3=y_1y_2-\beta y_3+u_3
\end{eqnarray}

where $u_1 $, $u_2 $, $u_3 $ are the controllers to be chosen later.
Here our aim is to determine the controller for the purpose of mixed synchronization with parameter mismatch.
Let the target state is defined as

\begin{equation}
\varphi (x) =\alpha x 
 \end{equation}
where $\alpha=(\alpha_{ij})_{n\times n}$ is the scaling matrix. For mixed synchronization we choose the scaling 
matrix as $\alpha=\mbox{diag}(\alpha_{11}, \alpha_{22}, \alpha_{33})$. When $\alpha_{ii}=1, i=1, 2, 3$, we can only achieve 
CS state without any amplification or attenuation. For $|\alpha_{ii}|>1 \;\;\mbox{or}\;\;  |\alpha_{ii}|<1, i=1, 2, 3$, scaling of the 
size (amplification or attenuation) of the driving attractor at the response attractor is also possible.  

The error system is defined by
\begin{eqnarray}
\label{error}
e_1=y_1-\alpha _{11}x_1\nonumber\\
e_2=y_2-\alpha _{22}x_2\nonumber\\
e_3=y_3-\alpha _{33}x_3
\end{eqnarray}

where $\alpha_{11},\alpha_{22},\alpha_{33}$ are the scaling factors.
Then the error dynamics can be written as
\begin{eqnarray}
\label{7}
&&\dot e_1=\sigma (y_2-y_1)+u_1-\alpha _{11}(\sigma +\Delta \sigma)(x_2-x_1)\nonumber\\
&&\;\;\;\;=\sigma (e_2-e_1)+\sigma \alpha _{22}x_2+u_1-\alpha _{11}\sigma x_2-\alpha _{11}\Delta \sigma (x_2-x_1)\nonumber\\
&&\;\;\;\;=\sigma (e_2-e_1)+\sigma x_2(\alpha _{22}-\alpha _{11})-\alpha _{11}\Delta \sigma (x_2-x_1)+u_1 \nonumber\\
&&\;\;\;\;=\sigma (e_2-e_1)+w_1
\end{eqnarray}

where $u_1=\alpha _{11}\Delta \sigma (x_2-x_1)-\sigma (\alpha _{22}-\alpha _{11})x_2+w_1$

\begin{eqnarray}
\label{7}
&&\dot e_2=ry_1-y_2-y_1y_3+u_2-\alpha _{22}(r+\Delta r)x_1+\alpha _{22}x_2+\nonumber\\
&&\;\;\;\;\;\;\;\alpha _{22}x_1x_3\nonumber\\
&&\;\;\;\;=r(y_1-\alpha _{11}x_1)+r\alpha _{11}x_1-e_2-y_1y_3+u_2\nonumber\\
&&\;\;\;\;\;\;\;-\alpha _{22}(r+\Delta r)x_1+\alpha _{22}x_1x_3\nonumber\\
&&\;\;\;\;=re_1-e_2+w_2
\end{eqnarray}

where $u_2=y_1y_3-\alpha _{22}x_1x_3+\alpha _{22}(r+\Delta r)x_1-r\alpha _{11}x_1+w_2$

\begin{eqnarray}
\label{8}
&&\dot e_3=y_1y_2-\beta y_3+u_3-\alpha _{33}x_1x_2+\alpha _{33}(\beta +\Delta \beta )x_3\nonumber\\
&&\;\;\;\;=-\beta e_3+w_3
\end{eqnarray}

where $u_3=-y_1y_2+\alpha _{33}x_1x_2-\alpha _{33}\Delta \beta x_3+w_3$

Thus, the system $(12-14)$ to be controlled is a linear system with the control input function $w(t)=[w_1(t),w_2(t),w_3(t)]^T$
as functions of the error states. When system $(12-14)$ is stabilized by the feedback $w(t)$, the error will converge to zero
as $t\longrightarrow \infty$ which implies that the system (8) and (9) are globally synchronized. To achieve this goal,
we choose $w(t)$ such that,
\begin{equation}
[w_1(t),w_2(t),w_3(t)]^T=A[e_1(t),e_2(t),e_3(t)]^T
\end{equation}

where $A=(a_{ij})_{3\times 3}$ is any matrix. Different choices of the matrix $A$ are possible. We make a choice of the matrix $A$ is as follows:
\begin{equation}
 A= \left(
  \begin{array}{ccc}
    0 & -\sigma & 0 \\
    -r & 0 & 0 \\
    0 & 0 & 0 \\
  \end{array}
\right)
\end{equation}
Other choices of the matrix $A$ are always possible. We re-define the feedback control function as follows:

\begin{equation}
\left( \begin{array}{c}
w_1 \\
w_2\\
w_3
\end{array} \right) =\left( \begin{array}{ccc}
0&-\sigma &0 \\
-r & 0 & 0\\
0 & 0 & 0
\end{array} \right)\times \left( \begin{array}{c}
e_1 \\
e_2\\
e_3
\end{array} \right)
\end{equation}

\begin{figure}
\includegraphics[width=120mm,height=80mm]{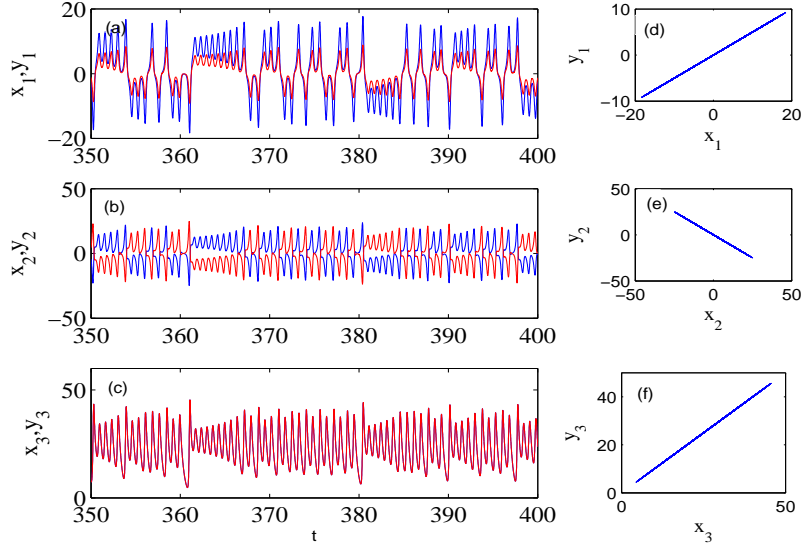}
\caption{\label . For mismatch Lorenz system $(a)$ time series of $ x_1$ in blue line and atunuated $y_1$ in red line with scaling factor $\alpha _{11}=0.5$
in CS state. (b) represents time series of $x_2$ in blue line and $y_2$ in red line with $\alpha _{22}=-1$ in AS state.  (c) time series of $ x_3$ in blue
line and response $y_3$ in red line in CS when $\alpha _{33}=1.$ and (d), (e), (f) are corresponding synchronization plot of (a), (b), (c) respectively. 
The parameters mismatch are $\bigtriangleup \sigma=1.0, \bigtriangleup r=0.0, \bigtriangleup \beta=0.0. $  }
\end{figure}

Hence the error system becomes
\begin{eqnarray}
\label{errlor}
&&\dot{e_1}=-\sigma e_1\nonumber\\
&&\dot{e_2}=-e_2 \nonumber\\
&&\dot{e_3}=-\beta e_3
\end{eqnarray}
Now we consider the Lyapunov function as follows:
\begin{equation}
V(t)=\frac{1}{2}(e_1^2+e_2^2+e_3^2)
\end{equation}
By using control law, the time derivative of $V$ is obtained as
\begin{eqnarray}
\label{vdotlor}
&& \dot{V}(t)=e_1\dot{e_1}+e_2\dot{e_2}+e_3\dot{e_3}\nonumber\\
&& \;\;\;\;\;\;\;\;=e_1(-\sigma e_1)+e_2(-e_2)+e_3(-\beta e_3)\nonumber\\
&& \;\;\;\;\;\;\;\;=-\sigma e_1^2-e_2^2-\beta e_3^2
\end{eqnarray}

Now from the Lyapunov stability function, we can say that the error system equation is asymptotically stable
if $\dot {V}\leq 0$, if $\sigma >0$ and $\beta>0$. We integrated numerically system (8) and (9) with the controllers $(u_1, u_2, u_3)$ using fifth order Runge-Kutta-Fehlberg method with integration step $\bigtriangleup t=0.001$ and taking random initial conditions.
Numerical results of MS state are shown in Fig. 1 for two
coupled mismatch Lorenz oscillators eqn. (8) and eqn. (9). The time series for three pair of variables ($x_1, y_1$),($x_2, y_2$) and
($x_3, y_3$) are plotted in Figs. 1(a), 1(b) and 1(c) respectively. Synchronization plot also shown in Figs.1(d), 1(e) and 1(f) where ($x_1,y_1$)
and ($x_3,y_3$)  are in CS state whereas ($x_2,y_2$) is in AS state. This MS state with amplification or attenuation is not possible under linear 
conventional diffusive coupling between two non-identical chaotic systems\cite{yclai}. This is contrast of LFC coupling with the diffusive coupling.



\section{Targeting Generalized Synchronization}

A generalized synchronization scenario was reported earlier using linear diffusive coupling between two nonidentical or large parameter 
mismatch oscillators. In GS scenario the functional relationship is usually unknown. To study linear generalized synchronization, we consider 
the three dimensional Sprott oscillator\cite{sprott}, described by the following sets of equations:
\begin{eqnarray}
\label{spott}
&&\dot{x_1}=x_1x_2-x_3 \nonumber\\
&&\dot{x_2}=x_1-x_2 \nonumber\\
&&\dot{x_3}=x_1+ax_3
\end{eqnarray}
where $x_1,x_2,x_3$ are the state variables and $a$ is the only parameter of the above system. The system exhibits chaos when
$a=0.3 $.
\par The system $(21)$ with the controllers $u_1,$ $u_2,$ $u_3 $ to be chosen as driving system
\begin{eqnarray}
\label{spottd}
&&\dot{x_1}=x_1x_2-x_3+u_1 \nonumber\\
&&\dot{x_2}=x_1-x_2+u_2 \nonumber\\
&&\dot{x_3}=x_1+ax_3+u_3
\end{eqnarray}
So the response system as follows
\begin{eqnarray}
\label{spottr}
&&\dot{y_1}=y_1y_2-y_3\nonumber\\
&&\dot{y_2}=y_1-y_2 \nonumber\\
&&\dot{y_3}=y_1+ay_3
\end{eqnarray}

\subsection{Targeting Linear Generalized Synchronization}

For linear generalized synchronization, we consider the target function as 
\begin{equation}
\varphi (x) =\left( \begin{array}{c}
\phi_1 \\
\phi_2\\
\phi_3
\end{array} \right)\
=\alpha x
 =\left( \begin{array}{ccc}
\alpha_{11}&\alpha_{12} &\alpha_{13} \\
\alpha_{21}&\alpha_{22} &\alpha_{23} \\
\alpha_{31}&\alpha_{32} &\alpha_{33}
\end{array} \right)\times \left( \begin{array}{c}
x_1 \\
x_2\\
x_3
\end{array} \right)\
\end{equation}

Now our aim is to determine the controller for the purpose of linear generalized synchronization.
Let the error vectors are defined as
\begin{eqnarray}
\label{error}
e_1=y_1-\alpha _{11}x_1-\alpha _{12}x_2-\alpha _{13}x_3\nonumber\\
e_2=y_2-\alpha _{21}x_1-\alpha _{22}x_2-\alpha _{23}x_3\nonumber\\
e_3=y_3-\alpha _{31}x_1-\alpha _{32}x_2-\alpha _{33}x_3
\end{eqnarray}
where $(\alpha_{ij})_{3\times 3}$, $(i, j=1, 2, 3)$ are the scaling matrix.

Then error dynamics can be written as,
\begin{eqnarray}
\label{7}
&&\dot e_1=y_1y_2-y_3+u_1-\alpha _{11}(x_1x_2-x_3)\nonumber\\
&&\;\;\;\;\;\;\;\;\;\;\;-\alpha _{12}(x_1-x_2)-\alpha _{13}(x_1+ax_3)\nonumber\\
&&\;\;\;\;=y_1y_2=u_1-\alpha _{11}x_1x_2-(y_3-\alpha _{31}x_1\nonumber\\
&&\;\;\;\;\;\;\;\;\;\;-\alpha _{32}x_2-\alpha _{33}x_3)-\alpha _{31}x_1-\alpha _{32}x_2\nonumber\\
&&\;\;\;\;\;-\alpha _{33}x_3+\alpha _{11}x_3-\alpha _{12}(x_1-x_2)-\alpha _{13}(x_1+ax_3)\nonumber\\
&&\;\;\;\;=-e_3+w_1
\end{eqnarray}
where $u_1=\alpha _{11}x_1x_2-y_1y_2+x_1(\alpha _{31}+\alpha _{12}+\alpha _{13})
+x_2(\alpha _{32}-\alpha _{12})+x_3(\alpha _{33}-\alpha _{11}+a\alpha _{13})+w_1$

\begin{eqnarray}
\label{10}
&&\dot e_2=y_1-y_2+u_2-\alpha _{21}(x_1x_2-x_3)-\alpha _{22}(x_1-x_2)\nonumber\\
&&\;\;\;\;\;\;\;\;-\alpha _{23}(x_1-ax_3)\nonumber\\
&&\;\;\;\;=e_1-e_2+\alpha _{11}x_1+\alpha _{12}x_2+\alpha _{13}x_3-\alpha _{21}x_1\nonumber\\
&&\;\;\;\;\;-\alpha _{22}x_2-\alpha _{23}x_3 +u_2-\alpha _{21}x_1x_2+\alpha _{21}x_3-\alpha _{22}x_1\nonumber\\
&&\;\;\;\;\;+\alpha _{22}x_2-\alpha _{23}x_1-a\alpha _{23}x_3\nonumber\\
&&\;\;\;\;=e_1-e_2+w_2
\end{eqnarray}

where $u_2=x_1(-\alpha _{11}+\alpha _{21}+\alpha _{22}+\alpha _{23})+x_2(-\alpha _{21})
+x_3(-\alpha _{13}+\alpha _{23}-\alpha _{21}+a\alpha _{23})+\alpha _{21}x_1x_2+w_2$

\begin{eqnarray}
\label{10}
&&\dot e_3=y_1+ay_3+u_3-\alpha _{31}(x_1x_2-x_3)-\alpha _{32}(x_1-x_2)\nonumber\\
&&-\alpha _{33}(x_1+ax_3)\nonumber\\
&&\;\;\;\;=e_1+ae_3+(\alpha _{11}x_1+\alpha _{12}x_2+\alpha _{13}x_3)+a(\alpha _{31}x_1\nonumber\\
&&+\alpha _{32}x_2)+\alpha _{33}x_3+u_3-\alpha _{31}x_1x_2+\nonumber\\
&&\alpha _{31}x_3-\alpha _{32}(x_1-x_2)-\alpha _{33}(x_1+ax_3)\nonumber\\
&&\;\;\;\;=e_1+ae_3+w_3
\end{eqnarray}

where $u_3=x_1(-\alpha _{11}-a\alpha _{31}+\alpha _{31}+\alpha _{33})+x_2(-\alpha _{12}-a\alpha _{32}-\alpha _{32})
+x_3-2\alpha _{31}+\alpha _{31}x_1x_2+w_3$

\begin{figure}
\includegraphics[width=125mm,height=95mm]{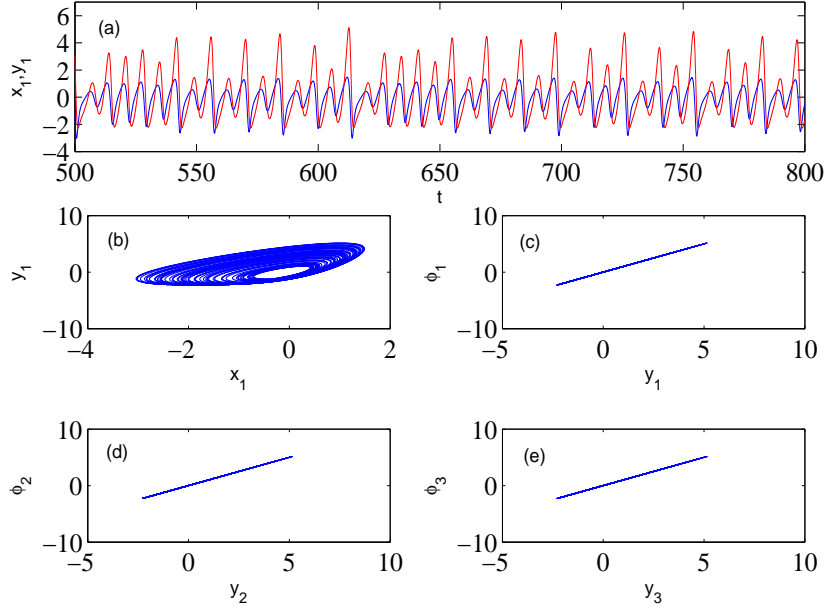}
\caption{\label . For Sprott system $(a)$ time series of $ x_1$ in blue line and $y_1$ in red line. (b) represents plot of $x_1$
with $y_1$.  (c) synchronization mainfold of ($ y_1,\phi_1$), (d) ($y_2,\phi_2$) and (e)($y_3,\phi_3$) where
$\phi_1=\alpha_{11}x_1+\alpha_{12}x_{2}+\alpha_{13}x_{3}$,
$\phi_2=\alpha_{21}x_1+\alpha_{22}x_{2}+\alpha_{23}x_{3}$,
$\phi_3=\alpha_{31}x_1+\alpha_{32}x_{2}+\alpha_{33}x_{3}$. All scaling parameters value are $\alpha_{ij}=1, \; i,j=1, 2, 3.$}
\end{figure}

Particular choice of linear feedback control function as,
 \begin{equation}
 \left( \begin{array}{c}
w_1 \\
w_2\\
w_3
\end{array} \right)
 =\left( \begin{array}{ccc}
-1&0&1\\
-1&0&0 \\
-1 &0& -2a
\end{array} \right)\times \left( \begin{array}{c}
e_1 \\
e_2\\
e_3
\end{array} \right)\
\end{equation}

Hence the error system become,
 \begin{eqnarray}
\label{errlinspott}
&&\dot e_1=-e_1\nonumber\\
&&\dot e_2= -e_2\nonumber\\
&&\dot e_3=-ae_3
\end{eqnarray}
The Lyapunov stability function

 \begin{eqnarray}
\label{errlinspott}
&&\dot V=\frac{1}{2}(e_1^2+e_2^2+e_3^2)=-(e_1^2+e_2^2+a e_3^2)
\end{eqnarray}

System is stable because $\dot V<0$. The time series, phase portrait, and synchronization plot of identical Sprott system is shown in Fig. 2.
In Fig. 2(a) shows time series of $x_1$ (blue color) and $y_1$ (red color) and which apparently shows no correlation by visual check. 
The synchronization plot in $(x_1, y_1)$ plane appeared to show no correlation. However, $y_1$ vs. $\phi_1$ in Fig. 2(c) shows clearly 1:1 
correlation and confirm that the targeting linear GS state. Other linear GS plot in $(y_2, \phi_2)$ and $(y_3, \phi_3)$ are shown in Figs. 2(d) and 2(e) respectively.

\subsection{Targeting Nonlinear Generalized Synchronization}
For nonlinear generalized synchronization we have taken a nonlinear target function such as,
\begin{equation}
\varphi (x) =\left(
\begin{array}{ccc}
 x_1^2\\
 x_2^2\\
 x_3
 \end{array}
\right)
\end{equation}
The synchronization errors become
\begin{eqnarray}
\label{error2}
e_1=y_1-x_1^2\nonumber\\
e_2=y_2-x_2^2\nonumber\\
e_3=y_3-x_3
\end{eqnarray}
So the error dynamics as follow
\begin{eqnarray}
\label{13}
&&\dot e_1=\dot y_1-2x_1(\dot x_1)\nonumber\\
&&\;\;\;\;=y_1y_2-y_3+u_1-2x_1(x_1x_2-x_3)\nonumber\\
&&\;\;\;\;=y_1y_2+u_1-2x_1^2x_2+2x_1x_3-(y_3-x_3)-x_3\nonumber\\
&&\;\;\;\;=-e_3+w_1
\end{eqnarray}
where $u_1=-y_1y_2+2x_1(x_1x_2-x_3)+x_3+w_1$

\begin{eqnarray}
\label{14}
&&\dot e_2=\dot y_2-2x_2(\dot x_2)\nonumber\\
&&\;\;\;\;=y_1-y_2+u_2-2x_2(x_1-x_2)\nonumber\\
&&\;\;\;\;=(y_1-x_1^2)+x_1^2-(y_2-x_2^2)+x_2^2-2x_1x_2+u_2\nonumber\\
&&\;\;\;\;=e_1-e_2+w_2
\end{eqnarray}

where $u_2=2x_1x_2-x_2^2-x_1^2+w_2$

\begin{eqnarray}
\label{14}
&&\dot e_3=\dot y_3-\dot x_3\nonumber\\
&&\;\;\;\;=y_1+ay_3+u_3+x_1-ax_3\nonumber\\
&&\;\;\;\;=(y_1-x_1^2)+a(y_3-x_3)+u_3+x_1^2-x_1\nonumber\\
&&\;\;\;\;=e_1+ae_3+w_3
\end{eqnarray}
where $u_3=-x_1^2+x_1+w_3$

The feedback controller for nonlinear generalized synchronization, we choose as 

\begin{equation}
 \left( \begin{array}{c}
w_1 \\
w_2\\
w_3
\end{array} \right)
 =\left( \begin{array}{ccc}
-1&0&1\\
-1&0&0 \\
1 &0& -2a
\end{array} \right)\times \left( \begin{array}{c}
e_1 \\
e_2\\
e_3
\end{array} \right)\
\end{equation}

Hence the error system become,
 \begin{eqnarray}
\label{errlinspottnon}
&&\dot e_1=-e_1\nonumber\\
&&\dot e_2= -e_2\nonumber\\
&&\dot e_3=-ae_3
\end{eqnarray}
The Lyapunov stability function

 \begin{eqnarray}
\label{errlinspott}
&&\dot V=\frac{1}{2}(e_1^2+e_2^2+e_3^2)=-(e_1^2+e_2^2+a e_3^2)<0
\end{eqnarray}
\begin{figure}
\includegraphics[width=125mm,height=85mm]{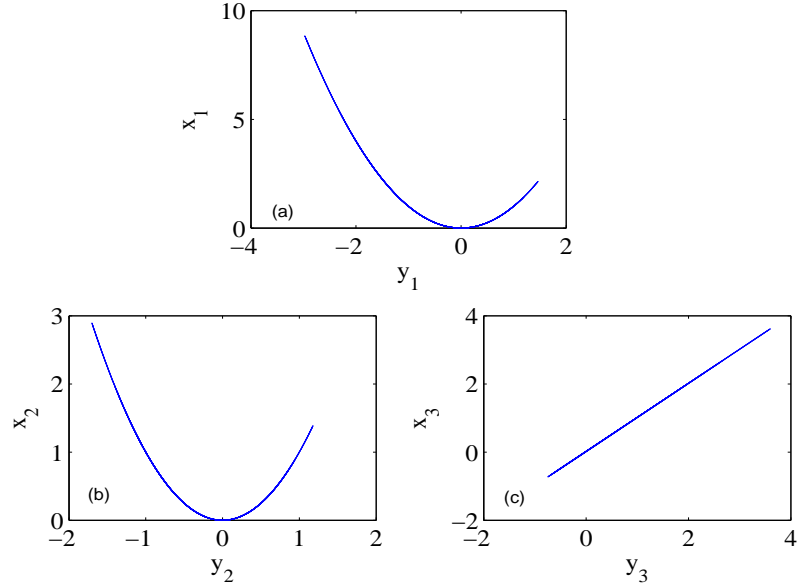}
\caption{\label . For Sprott system  nonlinear generalized synchronization plot of (a) $ (y_1, x_1)$, (b) $(y_2, x_2)$ and (c) $(y_3, x_3)$ 
where $\alpha_{11}=\alpha_{22}=\alpha_{33}=1.0$.}
\end{figure}

Fig. 3 shows nonlinear generalized synchronization plot. Figs. 3(a), 3(b) and 3(c) shown synchronization 
plot in $(y_1, x_1)$, $(y_2, x_2)$ and ($y_3, x_3$) respectively. By checking it is confirmed the onset of GS and satisfying 
the relations $y_1=x_1^2$, $y_2=x_2^2$ and $y_3=x_3$.

\section{Targeting fixed point}
For targeting fixed point of response system  the earlier Sprott system(21) is considered and target function as 
\begin{equation}
\varphi (x) = \alpha= \left(
\begin{array}{ccc}
 \alpha _{11} & 0 & 0\\
 0  &\alpha _{22} & 0\\
 0 & 0 & \alpha _{33}
 \end{array}
\right)
\end{equation}
The error sytem comes
\begin{eqnarray}
\label{error2}
e_1=y_1-\alpha _{11}\nonumber\\
e_2=y_2-\alpha _{22}\nonumber\\
e_3=y_3-\alpha _{33}
\end{eqnarray}

So the error dynamics as follows
\begin{eqnarray}
\label{14}
&&\dot e_1=\dot y_1=y_1y_2-y_3+u_1\nonumber\\
&&\;\;\;\;=y_1y_2-(y_3-\alpha _{33})-y_3\alpha _{33}+u_1\nonumber\\
&&\;\;\;\;=-e_3+w_1
\end{eqnarray}
where $u_1=\alpha _{33}y_3-y_1y_2+w_1$
\begin{eqnarray}
\label{15}
&&\dot e_2=\dot y_2=y_1-y_2+u_2\nonumber\\
&&\;\;\;\;=(y_1-\alpha _{11})+u_2+\alpha _{11}-\alpha _{22}\nonumber\\
&&\;\;\;\;=e_1-e_2+w_2
\end{eqnarray}
where $u_2=-\alpha _{11}+\alpha _{22}+w_2$
\begin{eqnarray}
\label{15}
&&\dot e_3=\dot y_3=y_1+ay_2+u_3\nonumber\\
&&\;\;\;\;=(y_1-\alpha _{11})+a(y_3-\alpha _{33})+u_3+\alpha _{11}+a\alpha _{22}\nonumber\\
&&\;\;\;\;=e_1+ae_3+w_3
\end{eqnarray}
 where $u_3=-\alpha _{11}+a\alpha _{33}$.
We choose the linear feedback function as 
 \begin{equation}
 \left( \begin{array}{c}
w_1 \\
w_2\\
w_3
\end{array} \right)
 =\left( \begin{array}{ccc}
-1&0&0\\
-1&0&0 \\
0 &0& -2a
\end{array} \right)\times \left( \begin{array}{c}
e_1 \\
e_2\\
e_3
\end{array} \right)\
\end{equation}
Finally the error system becomes
\begin{eqnarray}
\label{errlinspottnon}
&&\dot e_1=-e_1-e_3\nonumber\\
&&\dot e_2= -e_2\nonumber\\
&&\dot e_3=e_1-ae_3
\end{eqnarray}
So $e_1=e_2=e_3=0$ is asymptotically stable using Lyapunov stability theory.

It is also interesting that AD can be induced in the response system by taking $\alpha_{11}=-1.0, \alpha_{22}=0.1, \alpha_{33}=0.5$  as 
shown in Fig. 4. The driver $x(t)$ is oscillatory while the response $y(t)$ ceases to oscillate and is stable at targeted state. The coupling 
actually stabilizes the response system to any desired state. For $\alpha_{11}=\alpha_{22}=\alpha_{33}=0,$ the response system stabilized at origin (figures not shown). If the response system has no equilibrium point at origin, the coupling has
an inherent property to create a new equilibrium at origin and stabilize the response system at origin. For $\alpha_{11}=-1.0, \alpha_{22}=0.1,$ and $\alpha_{33}=0.5,$ the response system stabilize at $y_1=-1.0, y_2=0.1, y_3=0.5$ are shown in Fig. 4(a), 4(b), and 4(c) respectively plotted by solid red color lines. Whereas the driver states $x_1(t), x_2(t)$ and $x_3(t)$ are in chaotic state in Fig. 4(a), 4(b), 4(c) plotted by solid blue color lines. This is one of the major advantages
of this coupling over the conventional diffusive coupling. This targeting fixed point is used in practical system where chaos is an unwanted event.
\begin{figure}
\includegraphics[width=125mm,height=85mm]{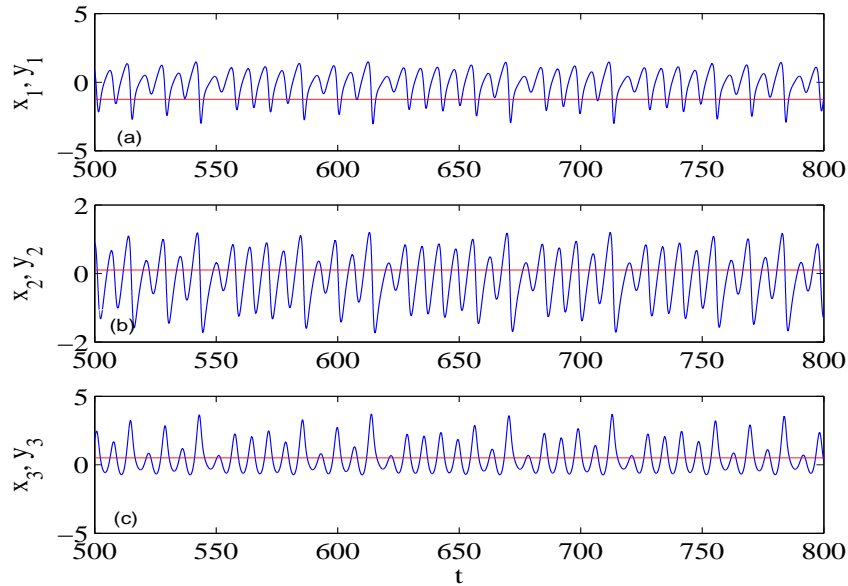}
\caption{\label . Time series of (a) $(x_1, y_1)$, (b) $(x_2, y_2)$ and (c) $(x_3, y_3)$ shows the driving system 
states $(x_1, x_2, x_3)$ (blue color) are in chaotic state and response system states $(y_1, y_2, y_3)$ (red color) are in 
targeted fixed point state. Here $\alpha_{11}=-1.0, \alpha_{22}=0.1, \alpha_{33}=0.5.$}
\end{figure}

\section{Conclusion}
We explore  linear feedback controller based coupling design using Lyapunov function stability theory
for targeting engineering synchronization  in chaotic oscillators.
The general scheme for coupling design in an unidirectional couple oscillators is discussed. By introducing a scaling factor
in the definition of coupling that allows amplification or attenuation of one attractor relative to another.
We described the theoretical details of the method about how to design coupling and illustrated with numerical examples of Lorenz 
and Sprott  systems. Compared to previous coupling \cite{yclai,uni,uni2,repulsive,inhibitory,inhibitory2,excitatory,synaptic}, our linear feedback controller based coupling has the following advantages: (i) previously mixed synchronization is observed using diffusive coupling in a chaotic flow with partial symmetry. We remove this restriction for mixed synchronization. (ii) Smooth control from one synchronization state to another synchronization state is possible by changing the scaling matrix in the coupling definition. (iii) Linear feedback controller coupling is independent on system parameters. (iv) In targetting engineering generalized synchronization state, the functional relation between drive and response state is known. But in conventional diffusive coupling, this functional relation for generalized synchronization in unknown always. (v) Amplification or attenuation of response system's attractors is possible. The scaling of chaotic attractor increases sucurity in chaos cryptography. (vi) Targeting engineering linear and nonlinear generalized synchronization 
using linear feedback coupling satisfying Lyapunov function stability theory were not explored earlier.  Physical realization of engineering 
synchronization using electronic circuit is also a challenging problem, will be the future work.


\bibliography{aipsamp}
\end{document}